\documentclass[aps,prd,onecolumn,groupedaddress,showpacs,nofootinbib,amssymb]{revtex4}
\usepackage[dvips]{graphicx}
\usepackage{amssymb}
\usepackage{amsmath}
\usepackage{graphicx}
\usepackage{amsfonts}
\usepackage{bm}
\usepackage{latexsym,float}
\usepackage{epsfig}
\usepackage{epstopdf}
\usepackage{amssymb}
\usepackage{verbatim}
\usepackage{multirow}
\usepackage{color}

\begin{document}

\title{Cosmology in modified $f(R,T)$-gravity}

\author{Petr V. Tretyakov$^{1,2}$}
\email{tpv@theor.jinr.ru} \affiliation{$^{1)}$Joint Institute for
Nuclear Research, Joliot-Curie 6, 141980 Dubna, Moscow region,
Russia\\ $^{2)}$Institute of Physics, Kazan Federal University,
Kremlevskaya street 18, 420008 Kazan, Russia}

\begin{abstract}
In present paper we propose further modification of $f(R,T)$-gravity (where $T$ is trace of energy-momentum tensor) by introducing higher derivatives matter fields. We discuss stability conditions in proposed theory and find restrictions for parameters to prevent appearance of main type of instabilities, such as ghost-like and tachyon-like instabilities. We derive cosmological equations for a few representations of theory and discuss main differences with convenient $f(R,T)$-gravity without higher derivatives. It is demonstrated that in presented theory inflationary scenarios appears quite naturally even in the dust-filled Universe without any additional matter sources. Finally we construct inflationary model in one of the simplest representation of the theory, calculate main inflationary parameters and find that it may be in quite agreement with observations.
\end{abstract}


\maketitle

\section{Introduction}\label{sec:1}

According to modern knowledge, based on experimental data, there was (at least) two different epochs of dynamical evolution of our Universe when the key role played dark energy (DE): inflationary stage at the early times of evolution and late time acceleration (l.t.a.) stage, which start recently (on cosmological scales) and continue till modern time. We know about existence of modern DE (associated with l.t.a.) with high precision from the different experiments first of which relate to SNI data \cite{Riess}, whereas about primordial DE (associated with inflation) we know only by indirect detections such as general isotropy and flatness of observable part of Universe and non-flatness spectrum of primordial scalar perturbations \cite{Planck1,Planck2}. Nevertheless the true nature of both DEs is unknown yet and this fact stimulates researchers to find solution of DE problem outside of standard physic.

Modifications of gravitational sector well known from early times and still very popular, because different corrections to gravitational action follows for instance from string theory \cite{Tseytlin} and one-loop quantum effects \cite{Birrell}(see also \cite{Starobinsky} for cosmological applications). The number of different approaches on this way actually huge and we only mention here such as $f(R)$-gravity \cite{Odintsov1}, Horndeski theory \cite{Kobayashi}, unimodular gravity \cite{Odintsov2}, teleparallel gravity \cite{AP}, theories with non-minimal kinetic coupling \cite{Sushkov}, and so on \cite{CFPS}.

Nevertheless there is another possibility to solve DE problem: we can introduce some exotic matter or modify right hand side (matter sector) of equations. The activity in this direction is not so intensive, but we can mention such attempts as phenomenological higher derivative matter fields \cite{Pani}, bulk viscosity and imperfect fluids \cite{Odintsov3}, theories with non-minimally coupled Ricci scalar with matter lagrangian \cite{NOP} and one of the most popular subclasses of this model $f(R,T)$-gravity \cite{Odintsov4}, where $T$ is the trace of energy-momentum tensor (stress-energy tensor). Note that dependence from $T$ may be induced by exotic imperfect fluids or quantum effects (such as conformal anomaly). Also we can study such kind of models as some phenomenological models, which arise from some more general theories. Indeed it is well known that brane models can modify exactly r.h.s. of equations of motions on the brane \cite{RS,DGP}. And by this reasons in our paper we try to discuss more wide class of $f(R,T)$-gravity and incorporate in function dependence from derivatives of $T$ (models contained $\Box R$-terms also are known as possible modification of $f(R)$-gravity \cite{Ovrut}).

This paper is organized as follows: in sec.\ref{sec:2} we derive general equations and discuss stability conditions; in sec.\ref{sec:3} we study a few concrete examples of functions and find some cosmological solutions;  in sec.\ref{sec:4} we estimate inflationary parameters for one of the simplest shape of function; and in sec.\ref{sec:5} we give some concluding remarks.

\section{General equations and stability conditions.}\label{sec:2}
Let us try to generalize well known modified gravity theory \cite{Odintsov4} by the next way

\begin{equation}
S=\frac{1}{16\pi}\int d^4 x \sqrt{-g} F(R,\,T,\Box T)+\epsilon\int d^4x\sqrt{-g}L_m,
 \label{1.1}
\end{equation}
where $R$ is Ricci scalar and $T$ is the trace of energy-momentum tensor and $\epsilon$ is equal to $1$ or to $0$. First of all let us ensure that this theory is ghost-free. For this task let us introduce Lagrange multipliers by the next way\footnote{In this section we exclude from discussion $L_m$ and concentrate our attention on function $F$.}

\begin{equation}
S=\int d^4 x \sqrt{-g} F(R,\,T,\Box T)=\int d^4x\sqrt{-g} [F(\lambda_1,\lambda_2,\lambda_3)+\mu_1(R-\lambda_1) +\mu_2(T-\lambda_2)+\mu_3(\Box T -\lambda_3)],
 \label{1.2}
\end{equation}
variations with respect to $\mu_i$ give us
\begin{equation}
\lambda_1=R,\,\,\,\lambda_2=T,\,\,\,\lambda_3=\Box T,
 \label{1.3}
\end{equation}
and variation with respect to $\lambda_i$:

\begin{equation}
\mu_1=F_{\lambda_1},\,\,\,\mu_2=F_{\lambda_2},\,\,\,\mu_3=F_{\lambda_3},
 \label{1.4}
\end{equation}
thus initial action (\ref{1.1}) may be rewritten in the form
\begin{equation}
S=\int d^4x\sqrt{-g} [ \mu_1R +\mu_3\Box\lambda_2 + \{ F(\lambda_1,\lambda_2,\lambda_3)-\mu_1\lambda_1 -\mu_3\lambda_3\}].
 \label{1.5}
\end{equation}
We can see that field $\mu_2$ is non-physical and as consequence decouple from equation. Now let us focus no the second term. It may be reorganized by introducing new fields $\lambda_2=\chi_2+\psi_2$ and $\mu_3=\chi_2-\psi_2$:
\begin{equation}
S=\int d^4x\sqrt{-g}\mu_3\Box\lambda_2 = \int d^4x\sqrt{-g}[\chi_2\Box\chi_2 -\psi_2\Box\chi_2 +\chi_2\Box\psi_2 -\psi_2\Box\psi_2 ]= \int d^4x\sqrt{-g}[-\nabla^i\chi_2\nabla_i\chi_2 +\nabla^i\psi_2\nabla_i\psi_2 ],
 \label{1.6}
\end{equation}
where we used integration by the parts. Formula (\ref{1.6}) tell us that independently on the type of function $F(R,\,T,\Box T)$ this theory contain three scalar fields (one additional will appear after $\mu_1 R$ decoupling) and at least one from it is ghost-like. But there is one special case\footnote{It is need to note that more general case of infinite rows like $\sum c_i\Box^iT$ may produce ghost-free theory, as it was happen in pure $f(R)$-gravity case \cite{BMS}.}, which allow us to solve this problem: if we put

\begin{equation}
F(R,\,T,\Box T)= f(R,\,T) + h(T)\Box T .
 \label{1.7}
\end{equation}
Note that $h(T)\neq\mathrm{ const}$ because in this case contribution to the equations will trivial.
In this case theory will contain only two physical scalar fields and both of them may be non-ghost depending on signs of $h'$ and $f_R$. Indeed, if we start from lagrangian (\ref{1.7}) and introduce auxiliary fields as $\lambda=R,\,\,\,\mu=f_{\lambda}$ we gain the next action
\begin{equation}
S=\int d^4x\sqrt{-g} [ \mu R + V(\mu,\lambda,T) - h' g^{ik}\nabla_i T\nabla_k T],
 \label{1.8}
\end{equation}
where potential $V=f(\lambda, T)-\mu\lambda$ and last term from (\ref{1.7}) was integrated by the parts. Further, producing conformal transformation of the metric $\bar{g}_{ik}=e^{\chi}g_{ik}$, $\chi=\ln\mu$ we have the action in canonical form
\begin{equation}
S=\int d^4x\sqrt{-\bar{g}} [  \bar{R} -\frac{3}{2}\bar{g}^{ik}\bar{\nabla}_i\chi\bar{\nabla}_k\chi +e^{-2\chi} V(\mu,\lambda,T) - h' e^{-\chi} \bar{g}^{ik}\bar{\nabla}_i T\bar{\nabla}_k T].
 \label{1.9}
\end{equation}
We can see that last kinetic term contain multiplier $h'/f_R$ thus we need $f_R>0$ and $h'>0$ for ghost-free theory.

Now varying lagrangian (\ref{1.7}) with respect to $T$ we find field equation
\begin{equation}
h' \Box T + \Box h+f_T(R,T)=2h' \Box T + h''(\nabla^i T)(\nabla_iT)+f_T(R,T)=0,
 \label{1.10}
\end{equation}
and finally varying (\ref{1.7}) with respect to metric we have
\begin{equation}
f_RR_{ik}-\frac{1}{2}Fg_{ik}+(g_{ik}\Box -\nabla_i\nabla_k)f_R= 8\pi\epsilon T_{ik} - (f_T+h'\Box T+\Box h)(T_{ik}+\Theta_{ik})+h'\nabla_iT\nabla_kT-\frac{1}{2}h'\nabla_mT\nabla^mTg_{ik},
 \label{1.11}
\end{equation}
where
\begin{equation}
\Theta_{ik} \equiv g^{lm}\frac{\delta T_{lm}}{\delta g^{ik}}.
 \label{1.12}
\end{equation}
We can see that if take into account field equation (\ref{1.10}), Einstein-like equation (\ref{1.11}) has essential simplification
\begin{equation}
f_RR_{ik}-\frac{1}{2}Fg_{ik}+(g_{ik}\Box -\nabla_i\nabla_k)f_R= 8\pi\epsilon T_{ik}+h'\nabla_iT\nabla_kT-\frac{1}{2}h'\nabla_mT\nabla^mTg_{ik}.
 \label{1.11.1}
\end{equation}
 Let us consider solution of equation (\ref{1.10}) in the form $T=T_0+\delta T$ and for the trivial solution $R_0=0$, $T_0=0$ (under the flat background) we find additional restriction $f_{TT}\geqslant 0$ for the absence of tachyon-like effective particles in the theory (case $f_{TT}=0$ can not be totally excluded). For more complicate cases of non-flat background this relation will has more complicate structure and will contain $h''$ as well, but it is clear that theory may be free from tachyon instability.

Let us take divergence of equation (\ref{1.11}). Divergence of l.h.s. reads

\begin{equation}
\begin{array}{l}
\nabla^i\left[ f_RR_{ik}-\frac{1}{2}Fg_{ik}+(g_{ik}\Box -\nabla_i\nabla_k)f_R \right]= (\nabla^i f_R)R_{ik} + f_R\nabla^iR_{ik}-\frac{1}{2}f_R\nabla_kR-\frac{1}{2}f_T\nabla_kT + (\nabla_k\Box -\Box\nabla_k)f_R\\
\\
-\frac{1}{2}\nabla_k(h\Box T)=(\nabla^i f_R)R_{ik} + f_R\nabla^iG_{ik}-\frac{1}{2}f_T\nabla_kT - (\nabla^i f_R)R_{ik}-\frac{1}{2}\nabla_k(h\Box T)=-\frac{1}{2}f_T\nabla_kT-\frac{1}{2}\nabla_k(h\Box T),
\end{array}
 \label{1.12.1}
\end{equation}
where we following to \cite{Ko1} used $\nabla^iG_{ik}=0$ and $(\nabla^i f_R)R_{ik}=(\Box\nabla_k-\nabla_k\Box )f_R$.
Thus from r.h.s. conservation equation now reads

\begin{equation}
\begin{array}{r}
(8\pi\epsilon-f_T-h'\Box T-\Box h)\nabla^iT_{ik}=(T_{ik}+\Theta_{ik})\nabla^i(f_T+h'\Box T+\Box h)+(f_T+h'\Box T+\Box h)\nabla^i\Theta_{ik}-\frac{1}{2}f_T\nabla_kT\\
\\
-\frac{1}{2}\nabla_k(h\Box T)-\nabla^i\left [h'\nabla_i T\nabla_k T \right ] +\frac{1}{2}\nabla_k(h'\nabla_lT\nabla^lT) ,
\end{array}
 \label{1.13}
\end{equation}
which is also may be simplified by using (\ref{1.10}) as
\begin{equation}
8\pi\epsilon\nabla^iT_{ik}=-\frac{1}{2}f_T\nabla_kT
-\frac{1}{2}\nabla_k(h\Box T)-\nabla^i\left [h'\nabla_i T\nabla_k T \right ] +\frac{1}{2}\nabla_k(h'\nabla_lT\nabla^lT) .
 \label{1.13.1}
\end{equation}

Note here very essential thing. Equations (\ref{1.11.1}) and (\ref{1.13.1}) are not contain true limits at $h=0$. If we want to find this limits, we must use equations (\ref{1.11}) and (\ref{1.13}). The reason of such kind situation quite understandable: for the limit $h=0$ field equation (\ref{1.10}) just absent (trivial) and simplifications which was produced became impossible. Exactly by this reason the proposed theory has very significant difference with respect to usual $F(R,T)$-gravity.

\section{Some concrete examples for cosmological applications.}\label{sec:3}

Now let us discuss some particular cases of gravitational field equations.
For cosmological application usually used
\begin{equation}
T_{ik}=(\rho+p)u_iu_k-pg_{ik},
 \label{2.1}
\end{equation}
with $u_iu^i=1$ and $u^i\nabla_ku_i=0$. In this case expression for $\Theta_{ik}$ takes very simple form
\begin{equation}
\Theta_{ik}=-2T_{ik}-pg_{ik}.
 \label{2.2}
\end{equation}

\subsection{The simplest case of functions: $f(R,T)=R+2f(T)$, $h(T)=\alpha T$.}\label{sec:3.1}

Let us discuss firstly the simplest case $f(R,T)=R+2f(T)$ for the universe with FLRW-metric
\begin{equation}
ds^2=dt^2-a^2(t)(dx^2+dy^2+dz^2),
 \label{2.1}
\end{equation}
filled by the dust matter ($p=0$, $T=\rho$). In this case equation (\ref{1.11}) gives us

\begin{equation}
3\frac{\dot a^2}{a^2}=3H^2=8\pi\epsilon\rho+f(\rho)+\frac{1}{2}\alpha\dot\rho^2+\frac{1}{2}\alpha\rho(\ddot\rho+3H\dot\rho),
 \label{2.2}
\end{equation}
\begin{equation}
2\frac{\ddot a}{a}+\frac{\dot a^2}{a^2}=2\dot H +3H^2=f(\rho)-\frac{1}{2}\alpha\dot\rho^2+\frac{1}{2}\alpha\rho(\ddot\rho+3H\dot\rho),
 \label{2.3}
\end{equation}
equation (\ref{1.10}) reads
\begin{equation}
\alpha(\ddot\rho+3H\dot\rho)=-f'(\rho),
 \label{2.4}
\end{equation}
and finally equation (\ref{1.13}) gives us
\begin{equation}
8\pi\epsilon(\dot\rho+3H\rho)= -f'(\rho)\dot\rho -\frac{1}{2}\alpha\rho\frac{d}{dt}\Box\rho-\frac{3}{2}\alpha\dot\rho\Box\rho,
 \label{2.5}
\end{equation}
which may be transformed to
\begin{equation}
8\pi\epsilon(\dot\rho+3H\rho)=\frac{1}{2}f'\dot\rho+\frac{1}{2}f''\rho\dot\rho,
 \label{2.6}
\end{equation}
It is easy to test our system: let us take time derivative from (\ref{2.2}), add (\ref{2.3}) multiplied by $-3H$ and cut from result $-9H^3$-term by using (\ref{2.2}), as result we gain equation (\ref{2.5}).

Now we can see that in this simplest case it is possible to write Fridman-like equation in the form $H^2=b(\rho)$, where $b$ -- some function of $\rho$. Indeed, expressing $\dot\rho$ from (\ref{2.6}) and substituting to (\ref{2.2}) and taking into account (\ref{2.4}) we find
\begin{equation}
3H^2\left [1-\frac{\alpha}{2}\frac{3(8\pi\epsilon\rho)^2}{(\frac{1}{2}f'(\rho)+\frac{1}{2}f''\rho-8\pi\epsilon)^2}\right ]=8\pi\epsilon\rho-\frac{1}{2}\rho f'+f,
 \label{2.7}
\end{equation}
which for instance for $f=2\lambda\rho$, $\epsilon=1$ reads
\begin{equation}
3H^2\left[ 1 -\frac{3\alpha}{2}\frac{(8\pi\rho)^2}{(\lambda-8\pi)^2} \right ]=(8\pi+\lambda)\rho,
 \label{2.8}
\end{equation}
and for $f=\lambda\rho^2$, $\epsilon=1$
\begin{equation}
3H^2\left[ 1 -\frac{3\alpha}{2}\frac{(8\pi\rho)^2}{(2\lambda\rho-8\pi)^2} \right ]=8\pi\rho.
 \label{2.9}
\end{equation}
Note that these expressions look like expressions, which arise from brane cosmological models.

Also it is useful calculate value $w_{eff}\equiv-1-2\dot H/H^2$:
\begin{equation}
w_{eff}=\frac{ -2f+\rho f'+\alpha\dot\rho^2 }{16\pi\rho\epsilon+2f -\rho f' +\alpha\dot\rho^2 }.
 \label{2.9.1}
\end{equation}

\subsubsection*{Comparison with the case $h=0$ and analogy with scalar field inflation.}

First of all note, that equations (\ref{2.2}) and (\ref{2.3}) very similar to equations described cosmology with scalar field (we need to substitute (\ref{2.4}) there). Indeed there is a kinetic term $\frac{1}{2}\alpha\dot\rho^2$ and some kind of potential $f-\frac{1}{2}\rho f'$. This fact very well understandable from (\ref{2.9.1}): if we can neglect by $16\pi\rho$ with respect to $2f -\rho f'$ (or put by the hand $\epsilon=0$) we obtain $w_{eff}=-1$ in slow-roll regime when $\dot\rho^2\ll 2f -\rho f'$. It means the classical inflation on the scalar field. Thus our new term has behavior absolutely identical to the scalar field one.

Now let us compare our theory with the limit case $h=0$, which was studied in previous investigations. In this case field equation (\ref{1.10}) is absent and we need to use (\ref{1.11}) instead of (\ref{1.11.1}). Equations (\ref{2.2}) and (\ref{2.3}) now reads

\begin{equation}
3H^2=8\pi\epsilon\rho+f+2\rho f',
 \label{2.9.2}
\end{equation}
\begin{equation}
2\dot H +3H^2=f,
 \label{2.9.3}
\end{equation}
which corresponds to eos
$$
w_{eff}=\frac{-f}{8\pi\epsilon\rho+f+2\rho f'},
$$
and not equal to $-1$ even for $\epsilon=0$. Moreover there is no kinetic term here, which may provide exit from inflation. Thus we can see that difference is very significant.

\subsubsection*{Unification of inflation and l.t.a.}

Note here also that it is possible to construct cosmological model which will unify inflation and late time acceleration by special shape of function $f(\rho)$. Indeed let us put
\begin{equation}
f(\rho)=\frac{a_1\rho^n+b_1\rho^m}{a_2\rho^n+b_2\rho^m},
 \label{2.9.4}
\end{equation}
where we imply all constants are positive and $n>m>0$. This function has the limits
\begin{equation}
\lim_{\rho\rightarrow+\infty} f(\rho)=\frac{a_1}{a_2},\,\,\,\lim_{\rho\rightarrow+0} f(\rho)=\frac{b_1}{b_2},
 \label{2.9.5}
\end{equation}
and we can see from equations (\ref{2.2}) and (\ref{2.3}) that term $f$ will play the role of cosmological constant in the beginning (case of big values of $\rho$ ) and in the end (case of small values of $\rho$) of Universe evolution, whereas existence of kinetic term $\dot\rho^2$ may provide transition between these two limit regimes. So in realistic cosmological models constants must satisfy next conditions:
\begin{equation}
\frac{a_1}{a_2}\approx\Lambda_{inf},\,\,\,\frac{b_1}{b_2}\approx\Lambda_{0},
 \label{2.9.6}
\end{equation}
where $\Lambda_{inf}$ -- value of cosmological constant in inflation epoch and $\Lambda_0$ -- value of cosmological constant in present time.

\subsection{Non-minimally coupling case: $f(R,T)=R+2\gamma R T + 2f(T)$.}\label{sec:3.2}
In this case we have

\begin{equation}
3H^2(1+2\gamma\rho)=8\pi\epsilon\rho+f(\rho)+\frac{1}{2}h'(\rho)\dot\rho^2+\frac{1}{2}h(\ddot\rho+3H\dot\rho)-6\gamma H\dot\rho ,
 \label{2.14}
\end{equation}
\begin{equation}
(2\dot H +3H^2)(1+2\gamma\rho)=f(\rho)-\frac{1}{2}h'(\rho)\dot\rho^2+\frac{1}{2}h(\ddot\rho+3H\dot\rho)-2\gamma (\ddot\rho+2H\dot\rho),
 \label{2.15}
\end{equation}
equation (\ref{1.10}) reads
\begin{equation}
h'(\rho)(\ddot\rho+3H\dot\rho)+\frac{1}{2}h''(\rho)\dot\rho^2=-f'(\rho)+6\gamma (\dot H +2H^2),
 \label{2.16}
\end{equation}
and finally equation (\ref{1.13.1}) gives us
\begin{equation}
 8\pi\epsilon(\dot\rho+3H\rho)=6\gamma\dot\rho(\dot H+2H^2)-f'\dot\rho-\frac{3}{2}h'\dot\rho(\ddot\rho+3H\dot\rho)-\frac{1}{2}h''\dot\rho^3-\frac{1}{2}h\frac{d}{dt}(\ddot\rho+3H\dot\rho),
 \label{2.18}
\end{equation}
It is easy to verify our system like in previous case: taking time derivative from (\ref{2.14}), adding (\ref{2.15}) multiplied by $-3H$ and cut from result $-9H^3$-term by using (\ref{2.14}), we gain equation (\ref{2.18}). Note also that energy conservation law (\ref{2.18}) takes very complicate form in this case.

\subsubsection*{Special solution.}

Now let us try to find some special solutions. We will find solution in the form
$$
\rho=\rho_0\ln [1+(t_c-t)],
$$
which near the the critical point $t=t_c$ may be decomposed as
$$
\rho=\rho_0\left [ (t_c-t) -\frac{1}{2}(t_c-t)^2+\frac{1}{3}(t_c-t)^3  \right ]=\rho_0 (t_c-t) ,
$$
and therefor
$$
\dot\rho=\rho_0[-1+(t_c-t)],\,\,\,\,\ddot\rho=\rho_0[-1+2(t_c-t)].
$$

Now let us suppose for the sakes of simplicity $h=\alpha\rho$, and $f(0)=0$, which is quite natural if we don't want to introduce cosmological constant by the hands. Substituting these expressions to (\ref{2.14}) and taking into account that $f=f'(0)\rho$ we find near critical point $t=t_c$ algebraical equation for $H$
\begin{equation}
3H^2(1+2\gamma\rho_0x)+H\left[ \frac{3}{2}\alpha\rho_0^2x-6\gamma\rho_0(1-x) \right] +\left[ \frac{3}{2}\alpha\rho_0^2x-8\pi\epsilon\rho_0x -f'(0)\rho_0x -\frac{1}{2}\alpha\rho_0^2 \right]=0,
 \label{2.18.1}
\end{equation}
where we denote $x\equiv (t_c-t)$. Equation (\ref{2.18.1}) has the only positive solution, which is reads (after linearization with respect to $x$) as
\begin{equation}
H=\gamma\rho_0\left (1 +\sqrt{1+\frac{\alpha}{6\gamma}} \right )+xF=H_0+xF,
 \label{2.18.2}
\end{equation}
where $F$ is the some function of parameters. Of cause, such kind of solution may not exist for arbitrary set of parameters, so let us ensure that solution which we found satisfy to all equations from the system (\ref{2.14})-(\ref{2.18}). First all note that all these equations will have finite part, so for verifying we can neglect by the all terms $\sim x$. Substituting our solution to (\ref{2.16}) we find
\begin{equation}
6\gamma\dot H_0= f'(0)-12\gamma H_0^2-\alpha\rho_0(1+3H_0),
 \label{2.18.3}
\end{equation}
where $\dot H_0$ and $H_0$ denotes finite part of $\dot H$ and $H$ correspondingly. Substituting expression (\ref{2.18.3}) to (\ref{2.18}) we find
\begin{equation}
8\pi\rho_0= \frac{1}{2}\alpha\rho_0^2(1+3H_0),
 \label{2.18.4}
\end{equation}
and this expression provide us relation between $\rho_0$ and parameters of the theory, which quite natural because $\rho_0$ is integration constant and must be determined from equations (constraint equation). Finally combining (\ref{2.18.3}) and (\ref{2.15}) we have
\begin{equation}
H_0^2-\frac{1}{3\gamma}f'(0) +\frac{\alpha}{3\gamma}\rho_0(1+3H_0)-\frac{1}{2}\alpha\rho_0^2+2\gamma\rho_0+4\gamma\rho_0H_0=0,
 \label{2.18.5}
\end{equation}
which may be transform by using (\ref{2.14}) to
\begin{equation}
f'(0)=\alpha\rho_0(1+3H_0)+6\gamma^2\rho_0+18\gamma\rho_0H_0-\alpha\rho_0^2,
 \label{2.18.6}
\end{equation}
which tell us that found solution exist only for some specific shape of function $f$\footnote{This situation may be changed if we will use more general function $h$.}. As a dry remnant we have future solution near which for zero energy density $\rho$ we have non-zero Hubble parameter $H$\footnote{For this solution we have non-zero $\dot H$ as well, thus it is not correspond to exact dS-solution.}.

\subsubsection*{Special case $\gamma=0,\,\,f(T)=0$.}

In this special case equations (\ref{2.14})-(\ref{2.15}) reads
\begin{equation}
3H^2=8\pi\epsilon\rho+\frac{1}{2}h'(\rho)\dot\rho^2+\frac{1}{2}h(\ddot\rho+3H\dot\rho),
 \label{2.19}
\end{equation}
\begin{equation}
(2\dot H +3H^2)=-\frac{1}{2}h'(\rho)\dot\rho^2+\frac{1}{2}h(\ddot\rho+3H\dot\rho),
 \label{2.20}
\end{equation}
equation (\ref{2.16}) tell us
\begin{equation}
h'(\rho)(\ddot\rho+3H\dot\rho)+\frac{1}{2}h''(\rho)\dot\rho^2=0,
 \label{2.21}
\end{equation}
and equation (\ref{2.18}) takes the form
\begin{equation}
 8\pi(\dot\rho+3H\rho)=-\frac{3}{2}h'\dot\rho(\ddot\rho+3H\dot\rho)-\frac{1}{2}h''\dot\rho^3-\frac{1}{2}h\frac{d}{dt}(\ddot\rho+3H\dot\rho),
 \label{2.22}
\end{equation}
and effective eos
\begin{align}
w_{eff}=\frac{h'\dot\rho^2+\frac{hh''}{2h'}\dot\rho^2}{16\pi\epsilon\rho+h'\dot\rho^2-\frac{hh''}{2h'}\dot\rho^2}.
 \label{2.23}
\end{align}
We can see that in the case when $\dot\rho^2\ll\rho$, which is quite natural for the power-law solutions $H\propto 1/t$, $w_{eff}=0$ and dust stage is realizing in the Universe. In some sense this is the limit case without potential, but only with kinetic term. Note also that this term may play the key role near Big Rip solution.

\subsubsection*{Future singularities in special case $\gamma=0,\,\,f(T)=0$.}

Let us discuss possibility of future singularities in our theory. Since we interested in future singularities which may appears due to new kinetic term, we discuss this question in the special case $\gamma=0$, $f(T)=0$. According to conventional classification \cite{NOT} there are four types of future singularities, which may be parameterized as follows (for more details see original paper).  If present Hubble parameter $H$ near singularity, which occur at the moment $t_s$ as
\begin{align}
H=H_0+h_0(t_s-t)^{-\beta},
 \label{2.24}
\end{align}
values of parameter $\beta\geqslant 1$ corresponds to the Type I of singularities, values $-1<\beta<0$ to the Type II, values $0<\beta<1$ to the Type III and $\beta<-1$ to the Type IV \cite{BNO}. Let us study possibility of realization for every type consistently.

Type I. Let us suppose $h\propto\rho^n$ and $\rho\propto (t_s-t)^{\alpha}$. Substituting these relations and (\ref{2.24}) to (\ref{2.19}) and taking into account (\ref{2.21}), we find that such kind of particular solution may be realized for $\alpha=(1-\beta)/(n+1)$ and $\beta>1$. Thus we can see that Type I of future singularity, which is also known as "Big Rip" may appears due to our new terms.

Type II. For this type of singularity we have $\dot H\gg H^2=H_0^2$ near the $t_s$ point. It mean that terms from r.h.s. of equation (\ref{2.20}) also much more than $H^2$ and the only possibility to satisfy equation (\ref{2.19}) is to put $2h'^2=h''h$, which lead to very specific shape of function $h=-1/(C_1\rho+C_2)$ and contradict to our basic requirements for stability. Thus we can see that this type of future singularity can not be realized due to our new terms.

Type III. In this case situation very similar to previous one: we have $\dot H\gg H^2\gg 1$ and realization of this type of singularity is impossible.

Type IV. In this case we have $H=H_0$, $\dot H=0$, while higher derivatives of $H$ diverge. From (\ref{2.20}) we can see that the only possibility to satisfy this equation is to have $\dot\rho=\mathrm{const}\neq 0$ near the point $t_s$. It imply the only possible solution $\rho=\Lambda_0+\rho_0(t_s-t)$, but even in this case we need in additional condition $8\pi\Lambda_0+h'\rho_0^2=0$ to have consistent system (\ref{2.19})-(\ref{2.20}). To satisfy this condition we need to put $h'<0$, which contradict to general stability condition, or to put $\Lambda_0<0$ which broke null energy condition. Thus we can see that this type of singularity also can not be realized due to our new terms.

Finally we can see that only Type I of future singularities may appears in our theory, but it is also quite clear that in the most general case of non-minimal coupling $f(R,T)$ any types of future singularities may appear due to non-trivial dependence of function from $R$, as it happens in usual $f(R)$-gravity. We address this question to future investigations.

\section{Basic inflationary model and its parameters.}\label{sec:4}

In this section let us try to calculate parameters of inflation, which may be constructed by using models previously described. Tensor-scalar ratio  $r$ and spectral index of primordial curvature perturbations $n_s$ may be expressed by using slow-roll parameters by the next way

$$
n_s=1-6\epsilon+2\eta,
$$

$$
r=16\epsilon,
$$
where the slow-roll indices being defined in terms of the Hubble rate as follows
$$
\epsilon=-\frac{\dot H}{H^2},
$$
$$
\eta=\epsilon-\frac{\ddot H}{2H\dot H}.
$$

As example of calculations let us take the model described in sec. \ref{sec:3.1} with arbitrary function $h(\rho)$.\footnote{The simplest case discussed above can not produce viable inflationary parameters.} Since during inflation stage we have slow-roll approximation, we can put the next relations $\ddot\rho\ll H\dot\rho$ and $\dot\rho^2\ll f$ and now equations (\ref{2.2})-(\ref{2.4}) take the form
\begin{equation}
3H^2=8\pi\rho+f-\frac{h}{2h'}f',
 \label{3.1}
\end{equation}
\begin{equation}
3H^2+2\dot H =f-\frac{h}{2h'}f',
 \label{3.2}
\end{equation}
\begin{equation}
\dot\rho =-\frac{f'}{3Hh'}.
 \label{3.2.1}
\end{equation}
It is clear that in slow-roll approximation Hubble parameter change slowly, so we have $\dot H\ll H^2$ and from comparison of equations (\ref{3.1}) and (\ref{3.2}) we can see that it equal to each other only if we can neglect by the $8\pi\rho$ term with respect to $f$. Thus such kind of regime may be realized for any $f\propto \rho^n$ with $n>1$, because in this regime we have big values of $\rho$. Finally instead of (\ref{3.1})-(\ref{3.2}) we have now
\begin{equation}
3H^2=f-\frac{h}{2h'}f'.
 \label{3.2.2}
\end{equation}
Expressions for $\dot H$ and $\ddot H$ which is also needed for slow-roll parameters may be calculated by consistent differentiating of expression (\ref{3.2.2}). Indeed we have for instance for $\dot H$

\begin{equation}
6\dot H=\frac{1}{H}\left ( f'-\frac{h'}{2h'}f'-\frac{h}{2h'}f''+\frac{hh''}{2h'^2}f' \right )\dot\rho=\frac{-f'}{3H^2h'}\left ( \frac{1}{2}f'-\frac{h}{2h'}f''+\frac{hh''}{2h'^2}f' \right ).
 \label{3.3}
\end{equation}

Now according to definition the number of e-folding before end of inflation $N_e=\ln\frac{a_e}{a_*}$, where $a_e$ scale factor related to the end of inflation and $a_*$ scale factor related to $N_e$ number. By using the definition of Hubble rate this formula may be transformed as follows
$$
N_e=\int^{t_e}_{t_*}H(t)dt=\int^{\rho_e}_{\rho_*}H(\rho)\frac{d\rho}{\dot\rho}=\int^{\rho_*}_{\rho_e}3H^2h'\frac{d\rho}{f'},
$$
where we used (\ref{3.2.1}), and finally we obtain
\begin{equation}
N_e=\int^{\rho_*}_{\rho_e}\frac{h'}{f'}\left( f-\frac{h}{2h'}f' \right)d\rho,
 \label{3.5}
\end{equation}
where $\rho_*\gg\rho_e$. Now let us focus on the case of power functions and put $f=\mu\rho^m$, $h=\nu\rho^n$. In this case expression (\ref{3.5}) may be easily integrated and we have
\begin{equation}
N_e=\frac{\nu (2n-m)}{2m(n+1)}\rho_*^{n+1}.
 \label{3.6}
\end{equation}
Expression (\ref{3.2.1}) takes the form
\begin{equation}
\dot\rho=\frac{-\mu m}{\nu n}\frac{\rho^{m-n}}{3H},
 \label{3.7}
\end{equation}
and for Hubble rate and its derivatives we have
$$
3H^2=\mu\rho^m\frac{2n-m}{2n},
$$

$$
\dot H=\frac{-m^2\mu}{6n\nu}\rho^{m-n-1},
$$

$$
\ddot H=\frac{m^3\mu^2(m-n-1)}{6n^2\nu^2}\frac{\rho^{2m-2n-2}}{3H},
$$
and finally collecting all terms we find for slow-roll parameters the next expressions
$$
\epsilon = \frac{m}{2(n+1)N_e},
$$

$$
\eta=\frac{2m-n-1}{2(n+1)N_e}=\frac{1}{N_e}\left (\frac{m}{n+1}-\frac{1}{2} \right ),
$$
where we take into account solution (\ref{3.6}).

Now let us take for example $n=m=2$. In this case we have for $N_e=50$, $r=0.107$, $n_s=0.9667$; and for $N_e=60$, $r=0.09$, $n_s=0.9778$. We can see that inflationary parameters lies near the boundary of viable region and taking more complicate functions may move it deeper in this region.

Finally let us ensure, that all variables has physical values. From (\ref{3.6}) we can see that $\rho_*$ has actually big value. $\dot\rho <0$ and $\dot H <0$ -- it means that both these variables are decrease during inflation as it must be. $\ddot H$ may has different sign depending on parameters, we can see that for $n=m=2$ it has negative values. And finally all derivatives $\dot\rho$, $\dot H$, $\ddot H$ must be small in comparison with $\rho$ and $H$ -- this fact put some additional restrictions for parameters $m$ and $n$ (otherwise our slow-roll approximation will broken). For instance in the case $n=m=2$ we have according to our formulas $\dot\rho\propto\rho^{-1}$, $\dot H\propto \rho^{-1}$ and $\ddot H\propto\rho^{-3}$ and since energy density $\rho$ has a big value all time derivatives actually small.

\section{Conclusions.} \label{sec:5}

In this paper we discuss possibility of further generalization of $f(R,T)$-gravity by incorporating higher derivative terms $\Box T$ in the action. First of all we find that in proposed theory inflationary scenarios appear quite naturally and may produce viable inflationary parameters. Moreover higher derivative terms decrease rapidly then the classic ones, but it may leads to future singularities of Type I. Another important thing: since new terms produce contribution to inflationary parameters it may resurrect such inflationary models as $R^n$ with $n>2$, which are already closed by modern observational data. We address this question for further investigations. It may be interesting also to generalize our theory by incorporating terms like $\sum c_i\Box^iT$, which may produce ghost-free theory for some specific sets of coefficients $c_i$. Thus we propose the theory, which is free from standard pathologies and hopeful for cosmological applications.

\begin{acknowledgments}

This work was supported by the Russian Science Foundation (RSF)
grant 16-12-10401.

\end{acknowledgments}

\end{document}